\newif\ifTECHRXIV
\TECHRXIVtrue
\ifTECHRXIV
    \documentclass{IEEEcsmagTechrxiv}
\else
    \documentclass{IEEEcsmag}
\fi

\newif\ifDEBUG
\DEBUGfalse

\newif\ifANONYMOUS
\ANONYMOUSfalse

\usepackage[colorlinks,urlcolor=blue,linkcolor=blue,citecolor=blue]{hyperref}
\expandafter\def\expandafter\UrlBreaks\expandafter{\UrlBreaks\do\/\do\*\do\-\do\~\do\'\do\"\do\-}
\usepackage{upmath,color}

\ifTECHRXIV
\else
    \jvol{XX}
    \jnum{XX}
    \paper{8}
    \jmonth{Month}
    \jname{IEEE Security \& Privacy Magazine}
    \jtitle{Special Issue on Secure Software Before Code/ing}
    \pubyear{2025}
\fi

\usepackage{amsmath,amsfonts}
\usepackage{algorithmic}
\usepackage{graphicx}
\usepackage{textcomp}
\usepackage{booktabs} % For formal tables \toprule
\usepackage{xspace} % Put an \xspace within numeric macros
\usepackage[normalem]{ulem}
\usepackage{makecell}
\usepackage{siunitx}
\usepackage[vskip=1em,font=itshape,leftmargin=2em,rightmargin=2em]{quoting}
\usepackage{listings}
\usepackage{subfigure}
\usepackage{url}
\def\UrlBreaks{\do\/\do-\do_} % Defining where it is okay to wrap a url to a new line. For bibliography.
\usepackage{soul}
\usepackage{hyperref}

\ifDEBUG
    \definecolor{olive}{rgb}{0.5, 0.5, 0.0}
    \definecolor{orange}{rgb}{1.0, 0.65, 0.0}
    \newcommand{\JD}[1]{\textcolor{blue}{[Jamie: #1]}}
    \newcommand{\TS}[1]{\textcolor{olive}{[Taylor: #1]}}
    \newcommand{\ST}[1]{\textcolor{red}{[Santiago: #1]}}
    \newcommand{\EB}[1]{\textcolor{orange}{[Ethan: #1]}}
    \newcommand{\KC}[1]{\textcolor{green}{[Kelechi: #1]}}

    \newcommand{\TODO}[1]{\hl{#1}}
    \newcommand{\CHOP}[1]{}

    \hypersetup{
        pdfpagemode=UseNone % don't show thumbnails or bookmarks on opening
    }

\else
    \newcommand{\JD}[1]{}
    \newcommand{\TS}[1]{}
    \newcommand{\ST}[1]{}
    \newcommand{\KC}[1]{}
    \newcommand{\EB}[1]{}
    \newcommand{\TODO}[1]{}
    \newcommand{\CHOP}[1]{}
    
\fi

\usepackage{balance} % Put \balance on the last page

\usepackage{cleveref}

\crefformat{section}{\S#2#1#3}
\crefrangeformat{section}{\S#3#1#4--\S#5#2#6}
\crefname{figure}{Figure}{Figures}
\crefname{appendix}{Appendix}{Appendices}
\crefname{table}{Table}{Tables}
\crefname{algorithm}{Algorithm}{Algorithms}
\crefname{listing}{Listing}{Listings}
\crefname{theorem}{Theorem}{Theorems}
\crefname{thm}{Theorem}{Theorems}
\crefname{lemma}{Lemma}{Lemmata}
\crefname{equation}{Eqt.}{Eqts.}
\crefformat{Grammar}{Grammar #1}

\newcommand{\ie}{\textit{i.e.,} }
\newcommand{\eg}{\textit{e.g.,} }
\newcommand{\etal}{\textit{et al.}\xspace}
\newcommand{\etals}{\textit{et al.'s}\xspace}

\newcommand{\myquote}[1]{\begin{quote}\emph{``#1''}\end{quote}}

\usepackage{quoting} % For customized quotation environments

\quotingsetup{
  leftmargin=0.25cm, % Adjust the left margin
  rightmargin=0.25cm, % Adjust the right margin
}

\newcommand{\firstlineindent}{\hspace*{0.3cm}}

\renewcommand{\myquote}[2]{
\begin{quoting}
{
\noindent
\small \firstlineindent \emph{``\textit{#1}'' \textnormal{--- #2}}
}
\end{quoting}}

\DeclareUnicodeCharacter{2060}{\nolinebreak}

\begin{document}

\ifTECHRXIV
\else
  \sptitle{Theme Article: Secure Software Before Codeing}
\fi

% \title{Signing: Current Practices and Future Trends}
\title{Establishing Provenance Before Coding: Traditional and Next-Gen Software Signing}

\author{Taylor R. Schorlemmer}
\affil{Purdue University, West Lafayette, IN, 47907, USA}

\author{Ethan H. Burmane}
\affil{Purdue University, West Lafayette, IN, 47907, USA}

\author{Kelechi G. Kalu}
\affil{Purdue University, West Lafayette, IN, 47907, USA}

\author{Santiago Torres-Arias}
\affil{Purdue University, West Lafayette, IN, 47907, USA}

\author{James C. Davis}
\affil{Purdue University, West Lafayette, IN, 47907, USA}

\ifTECHRXIV
\else
  \markboth{Secure Software Before Code/ing}{Secure Software Before Code/ing}
\fi

\begin{abstract}
	Software engineers integrate third-party components into their applications.
	The resulting software supply chain is vulnerable.
	To reduce the attack surface, we can verify the origin of components (provenance) before adding them.
    Cryptographic signatures enable this.
    This article describes traditional signing, its challenges, and the changes introduced by next-generation signing platforms.
    
\end{abstract}

\maketitle

%\keywords{Provenance; Software Signing; Software Re-Use; Component Selection; Cybersecurity; Software Engineering}

\begin{figure*}[!ht]
	\centering
	\includegraphics[width=1\linewidth]{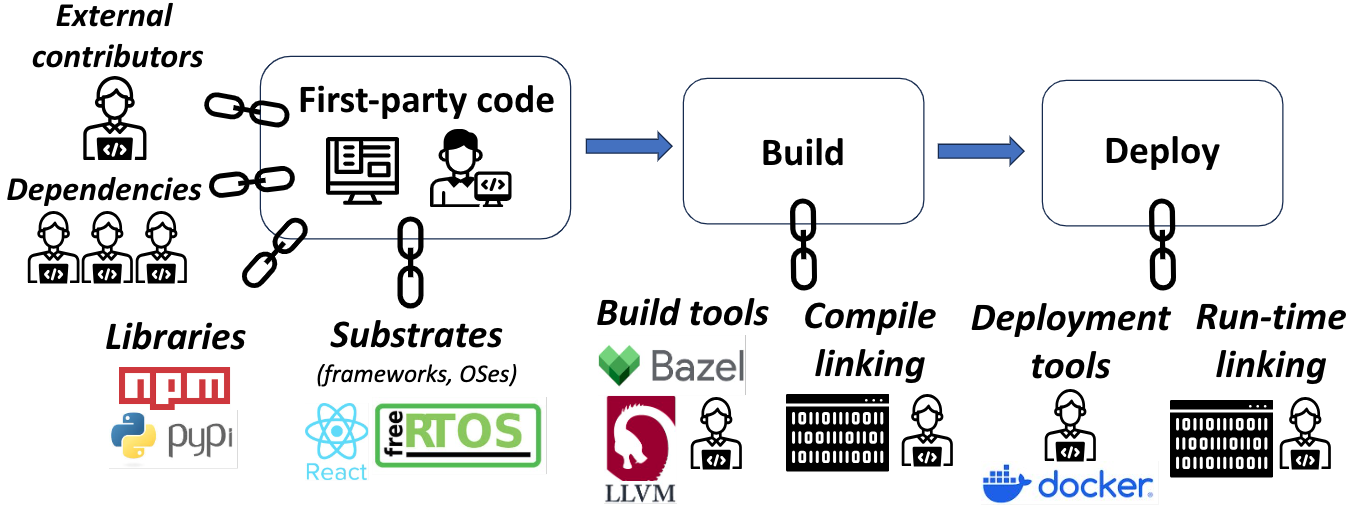}
	\caption{
		Depiction of the software factory model to highlight the external dependencies.
		Each chain symbol represents part of the software supply chain involved in producing software.
        Each component comes from some individual or organization.
        Provenance techniques enable downstream users to verify the origins of the components on which they depend.
	}
	\label{fig:ThirdPartyDecisionPoints}
\end{figure*}

% INTRODUCTION
% \JD{The connection between dependencies and `supply chain' should be made in this paragraph}
\chapteri{N}early all modern software relies on other software components~\cite{Sonatype2024}.
As illustrated in~\cref{fig:ThirdPartyDecisionPoints}, these components include libraries, operating systems, build tools, and deployment tools.
As part of the planning process, software engineers decide which components they will use to build their software.
These dependencies (and the components on which they depend in turn) create a \textit{software supply chain} with implied trust relationships between software components and their users.
When software engineers decide which components to use in their software, they are also deciding which components to trust.
%they choose to use third-party tools and dependencies to speed up development.
%Since these components can be developed by other teams or other organizations, the supply chain includes trust relationships between the product and its dependencies.

%Engineers must consider functional (fit for purpose) and non-functional properties when selecting components.
%These trust relationships went unacknowledged for a long time, but are now the focus of great interest.
Many recent cybersecurity attacks have exploited these trust relationships by targeting software components and supply chains~\cite{Sonatype2024}.
In response, many researchers have proposed enhancements software supply chain security.
Okafor \etals literature review summarized those proposals as addressing three distinct properties:
  separation to ensure that failures in one component are isolated,
  transparency to see the full supply chain,
  and
  validity to show that components are not changed unexpectedly (integrity)~\cite{okafor_sok}.
Taken together, these latter two properties can describe the \textit{provenance} of both an individual software component and of the resulting supply chain.

\iffalse
	Software supply chain security is a growing concern for software engineers who, after witnessing several high profile supply chain attacks, realize the need for a secure set of dependencies.
	This is a widespread concern for software engineers because most modern software systems rely on 3rd party dependencies.
	Engineers have a growing list of tools to secure their software supply chains, but one class of techniques - signing - is the de-facto method for assuring the origin of an artifact.
\fi

%This article focuses on these \textit{provenance techniques}, whicha class of techniques intended to increase transparency and validity: \textit{provenance}.
This article focuses on one specific technique for software component provenance: \textit{software signing}.
Software signing using public-key cryptography is the \textit{de facto} method for assuring the origin of an artifact.
Although signing is an old concept, there have been many works showing that traditional approaches to software signing is easier said than done.
However, next-generation signing platforms have emerged in recent years that improve usability.
%, and software signing is now commonplace by open-source and commercial vendors.

In this article, we summarize software supply chains, traditional methods for software signing, and the transformative aspects of next-generation software signing platforms.
We discuss the current state of software signing and describe how improvements to signing infrastructure are leading to better security practices.
%These developments are affecting component selection, because provenance information is now readily available for many packages.
Our goal is for the reader to learn that, due to the changing software signing landscape, their component selection process can, and should, integrate provenance information.

\myquote{A few years from now, everybody expects provenance...they don’t install anything if they can’t prove where it came from...Signing for open source is probably the single most important thing for software supply chain security}{Technical leader\footnote{This and other quotes are taken from interviews of industry cybersecurity experts conducted by Kalu \etal in 2025~\cite{KaluSigningARXIV}.}} % Kalu S4

\section{TRUST BEFORE CODING}

% In this section we discuss
% the role of component selection in software development,
% describe cybersecurity considerations for the resulting software supply chain,
% and indicate different approaches to establishing provenance in a software supply chain.

\subsection{Components Influence Product Code}

Before writing code, software engineers consider the components they will use to build their software.
%These components affect several aspects of the resulting software.
%This includes the security of the software, not just its functionality.
%For this reason, the choices made during component selection are critical.
% Understanding which components are used and where they come from is essential for securing the software supply chain.
%
%
%Even in a highly iterative (``agile'') software development shop, some decisions must be made after a product is conceptualized but before coding begins.
One of the primary decisions is this:
\textit{Which external components should we build on?}
\cref{fig:ThirdPartyDecisionPoints} depicts the kinds of components on which software depends, aligned with the Software Factory Model~\cite{SLSA_SFM}.
The first-party code in a software product may interact with external libraries --- \textit{Which ones}?
It may rely on a framework or an operating system for interaction with the outside world --- \textit{What is the resulting control flow, and which parts will be handled by the substrate?}
When being built --- \textit{Should it make use of Ant or Gradle or Bazel?}
When being deployed --- \textit{Should it rely on Docker, Podman, or Singularity as its virtualization scheme?}

Each of these decisions affects the resulting first-party code.
The selected libraries and substrates have an obvious effect: the code will call one API or another.
More subtle influences are exercised downstream.
Build tools may influence which language features are used, \eg if the product can make use of GCC extensions to C or must rely on the C90 standard.
Deployment tools may influence the parameterization of the system, \eg the size of buffers and caches (and which of these knobs to expose).

Since dependency selection is a weighty matter, engineers evaluate candidate components along many dimensions.
Functional considerations such as correctness and performance have long dominated this selection process.
However, the rise of cyberattacks has made cybersecurity, and particularly the security risks of depending on external components, a top-of-mind concern. % and the subject of government oversight (\eg EO-14028)~\cite{EO14028}.

\subsection{Provenance in Software Supply Chains}

Provenance is a security property that combines validity (data and actor integrity) and transparency.
Provenance simultaneously demonstrates that the code has not been modified, and gives evidence of the entity in control of the code at the time of release~\cite{okafor_sok, CooperNIST}.

In the past, assessing the provenance of a component was left to the judgment of the engineering team.
Now, the urgency of cyberattacks has driven greater top-down control to improve software supply chain security.
Security requirements related to component provenance are being introduced through industry standards and government regulations.
Relevant industry standards include
The Linux Foundation’s Supply Chain Levels for Software Artifacts (SLSA)~\cite{SLSA_SFM}, CNCF’s Software Supply Chain Best Practices,
%  the Open Web Application Security Project (OWASP),
  and Microsoft’s Secure Supply Chain Consumption Framework (S2C2F).
%  and Secure Supply Chain Consumption Framework (S2C2F).
The US government has also published standards such as NIST’s Secure Software Development Framework (NIST-SSDF) and NIST 800-204D, along with regulations for contractors handling government contracts (US Executive Order 14028~\cite{EO14028}).

While security techniques typically ensure distinct properties (\eg software attestations ensure transparency, software signing ensures validity), combining security techniques is often preferred in practice to reinforce one or more of these properties, with provenance being one such combination.
An example of a combination used to establish provenance involves signed attestations, which integrate validity (ensuring data and actor integrity) and transparency (identifying the entity in control at signing). The Software Bill of Materials (SBOM) is another method for establishing transparency that can be used in place of attestations.
Next, we briefly summarize these methods to establish provenance for software components.

\subsubsection{Software Signing}
Software signing is a method of verifying the origin of software artifacts.
% Users use digital signatures on source code, binaries, or even git commits to establish provenance.
Provenance techniques primarily rely on signing (\eg digital signatures on source code, binaries, or even git commits) to ensure the integrity of both the actor and the data within components.
Downstream users can verify the signature to ensure that the software artifact came from an expected source.
Like other provenance techniques, code signing does not make any quality guarantees about the software's correctness (it might still have bugs!), but it does provide a way to verify the origin of a piece of software.

%  Several techniques exist to establish provenance in software supply chains.
% Examples include software bills of materials (SBOMs), attestations, and code signing.
% Many of these techniques use some form of signing to ensure the integrity of the information they provide.
% This enables end users to verify that information has not been tampered with.

\subsubsection{Software Bills of Materials (SBOMs)}
SBOMs contain a list of all the components that make up a piece of software.
This list often includes the version of each component and the origin of the component.
SBOMs can be used by downstream users to understand the risks associated with a piece of software.
Furthermore, SBOMs can be used to ensure that a piece of software is built with the correct components.
SBOMs can be signed to prove that the manifest has not changed.

\subsubsection{Attestations}
Attestations are pieces of metadata that describe some aspect of a software artifact.
For examples, build attestations describe how a piece of software was built, and review attestations indicate that a particular person has reviewed a piece of code, a component, a process, etc.
Typically, these attestations are signed to ensure their integrity.
This enables users to verify that the metadata has not been tampered with.

\vspace{10pt}

In this article, we focus on the first of these techniques that underpin provenance --- \textit{software signing}.
Like all security practices, signing is not a silver bullet.
Although signing does not guarantee the correctness of a piece of software, it does provide a way to verify the origin of a piece of software and that it has not changed.
Most industry standards highlighted earlier emphasize the importance of signing, which is prominently featured in their recommendations.
For example, NIST 800-204D requires that attestations about software products (\eg their bills of materials/SBOMs, build processes, etc.) be cryptographically signed with a secure key.
NIST standards further require that consumers of software components verify these attestations before using them.
%Recently, several changes have been made to signing infrastructure that have improved the state of signing.
We review some of the shortcomings of traditional signing methods and how newer techniques are attempting to resolve these issues.
We first describe the foundations of software signing: Public-key cryptography.

\section{PRIMER ON PUBLIC-KEY CRYPTOGRAPHY}

Public-key cryptography, also known as asymmetric cryptography, refers to the subset of cryptographic methods that utilize a pair of public and private keys.
% The merits of these methods; scalable number of users, and no need of a secure channel enables several applications (\eg signing) where symmetric cryptography falls short.
% The presence of two keys enables several applications (\eg signing) where symmetric cryptography (single key) falls short.
The advantages of these methods, such as scalability for a large number of users and the elimination of the need for a secure channel, enable various applications (\eg signing) where symmetric cryptography falls short.
This section gives a brief primer --- see the References section for a starting point on this topic.

\subsection{Key Terms and Concepts}
In a public-key cryptographic system, an application or user can generate a pair of public and private keys.
The \textit{private key} is kept secret and the \textit{public key} is shared.
Data encrypted using one key can be decrypted using the other.
This leads to two basic use cases:
(1) holders of the public key encrypt messages that can only be read by holders of the private key; or
(2) holders of the private key sign messages (typically by appending an encrypted hash of the message) that can be verified by holders of the public key.
In either case, it is important to establish trust in the public key (\ie that the public key is from the correct source and has not been modified).
Methods such as public-key infrastructure and the web of trust have been used to strengthen this trust.

With respect to provenance, the second use case is especially important.
After establishing trust in the public key, signatures provide strong guarantees about the origin of a message.
This scheme is used by several computing applications, detailed next.

\subsection{Applications in Computing}
Public-key cryptography is widely used in computing.
For example, S/MIME (email communication), SSL/TLS (network communication), SSH (remote network access), and even smart cards use forms of public-key cryptology for secure communication and/or authentication.
Given the topic of this article, our primary interest is on the application of public-key cryptography to ensuring provenance in the software supply chain.
For this reason, we will focus on software-related signing tools (which also use public-key cryptography) such as Pretty Good Privacy (PGP) and Sigstore.
Let's take a look!

\section{TRADITIONAL SOFTWARE SIGNING}
\TS{This section focuses only on code signing (and so does the next gen one). We need to say that we're using software signing as an example here and in the status section. Other signing types exist.}
\KC{I added this info in the provenance section now}
Many platforms, such as software registries, support ``traditional'' signing as a way for engineers to verify the origin of the software components they rely on.
Traditional signing is relatively simple, but it has drawbacks.
When used correctly, traditional signing methods can cryptographically ensure the origin of an artifact.
Unfortunately, these methods suffer from poor adoption and key management issues that can limit practical use.

\subsection{How Traditional Signing Works}

\begin{figure*}[!ht]	\centering
	\includegraphics[width=1\linewidth]{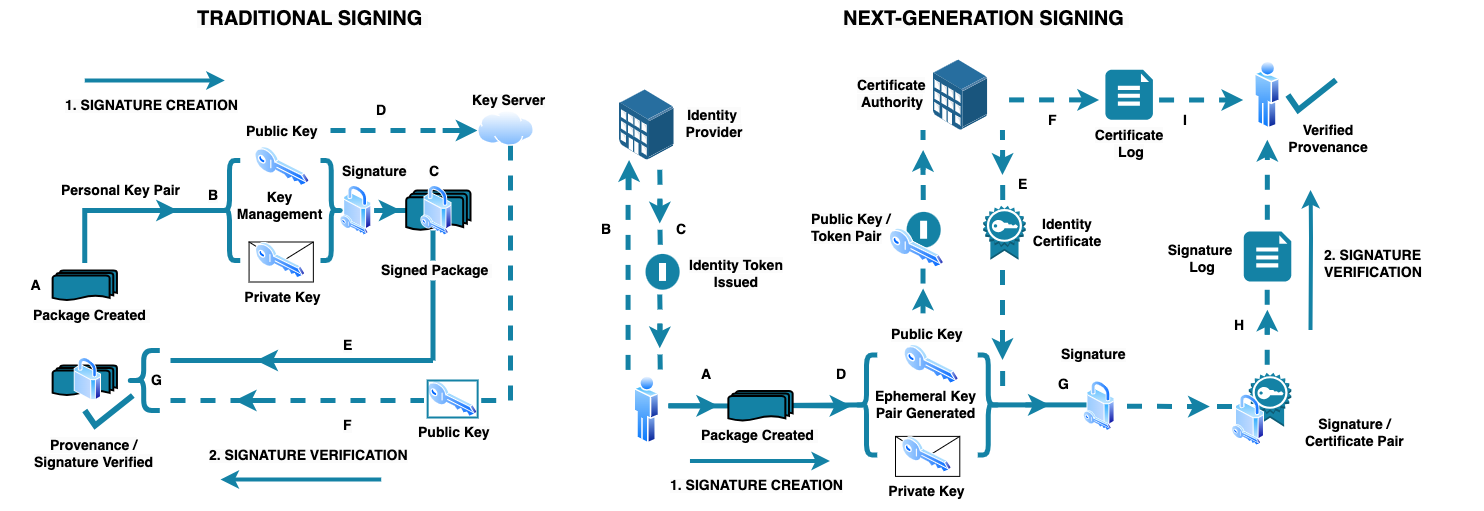}
	\caption{
	Traditional workflow for software signing (left) and next-gerneration workflow for software signing (right).
    Figure reproduced and adapted from Kalu \etal~\cite{KaluSigningARXIV}.
	}
	\label{fig:trad-vs-next-gen-signing}
\end{figure*}

Several implementations of traditional signing exist, but they share a common form.
The left half of \cref{fig:trad-vs-next-gen-signing} demonstrates how traditional signing is typically used to provide provenance for software~\cite{CooperNIST}. 
First, an author creates a package (A) and a public/private key pair (B).
Next, the author uses the private key to sign the package (C).
Then, the signed package (C) and the associated public key (D) are published.
To use a package, an engineer downloads the package (E) and the author's public key (F) before verifying the signature (G).
If an engineer trusts that the public key came from the author and the signature verifies successfully, they can be confident about the origin of the package.

\subsection{An Example: Pretty Good Privacy (PGP)}
There are several established software signing tools, such as PGP, Git commit signing, and Docker Content Trust.
These tools vary in the way that they implement the signing.
Pretty Good Privacy (PGP), is a popular and well established format standardized by OpenPGP.
It was originally created to secure email traffic, but has since been used to encrypt and sign a variety of other data types.
The OpenPGP standard is implemented in a few different ways, but a notable instance is GNU Privacy Guard (GPG).

Using GPG, artifact authors can create a key with \verb|gpg --gen-key|.
This walks them through creating a public/private key pair.
After generating a key pair, they can then sign some file with a command like \verb|gpg -ab some.file|.
This creates a detached signature file called \verb|some.file.asc|.
The author can then publish their file, signature, and public key.
After fetching the file, signature, and public key, an engineer downstream  can verify the file using \verb|gpg --verify some.file.asc some.file|.

% Traditional signing schemes rely on a simple cryptographic signing mechanism.
% That mechanism works as follows:
% An individual, lets call her Alice, has some sort of software artifact.
% This artifact might be a piece of code, some metadata, a binary, or really any piece of information.
% Someone else wants to use this artifact, lets call him Bob.
% If Bob wants to ensure that the artifact

\subsection{Challenges}
\myquote{A lot of the ways in which I think previous teams' software signing hasn't been useful is...key management...how the private key is managed but also in how the public key is made available}{Member of senior management} % Kalu S13

The example commands of GPG above seem simple.
In just two commands, a software producer can create keys and a signature, and a consumer can validate the signature in a single command.
However, there are several potential problems with this process.

Traditional signing methods have often been criticized for their lack of usability.
Academic literature, including works similar to the ``Johnny signs'' studies~\cite{JohnnySigns1}, has documented these criticisms.
In practice, there is a noticeable trend where several traditional signing tools are being discontinued by various software registries, such as PyPI's cessation of PGP use~\cite{schorlemmer_signing}.
Common usability issues are generally associated with key management, inadequate documentation, and user interface design challenges.

\subsubsection{Key Management}
Key management is particularly manual in traditional signing.
Users are responsible for securely storing their private key and sharing their public key.
This means that they must not lose the private key and must keep it private.
This becomes more complicated in organizational environments with personnel turnover.
When keys become compromised, or are otherwise no longer secure, all corresponding signatures loose trust.
This also necessitates the generation and distribution of new keys.

\subsubsection{Identity Verification}
Establishing trust in a public key is also particularly challenging.
Once a user creates a key pair, they must find a way to share their public key.
This is typically done through infrastructure like public key servers or through other file sharing means.
Unfortunately, many of these methods do not provide strong identity guarantees for the creator of the public key.
The web of trust was introduced as a way to introduce a more confident binding between an identity and a public key.
There are, however, even issues with this technique (\ie relying on social interactions like signing parties).

\subsubsection{Transparency}
Additionally, there are transparency issues with traditional signing schemes.
Users typically do not have knowledge about which artifacts were signed or what information should be present.
Without knowledge about what modifications have been made to a project's metadata, users cannot verify that supply chain actors are behaving properly.
For example, a downstream user may not be able to detect that the signature identity has changed in a project.

\subsection{Current State}
%\myquote{A lot of the ways in which I think previous...software signing hasn't been useful is...key management...in terms of how the private key is managed but also in how the public key is made available.}{Member of senior management} % Kalu S13
% Traditional signing has long been subject of criticism for its usability or lack thereof.
% Many "Johnny signs" type works document this in the academic literature. In practise we see this correlation in the several discontinuation of traditional signing tools from various software registries (\eg PypI discontinuing use of PGP), and organizations. Usability issues typically  revolve around Key management, documentation, and user interface design issues.

The current state of signing varies depending on the type and environment of the software being developed.
In general, signing is more prevalent in commercial software than in open source software.
This indicates that provenance is lacking in some open source environments.

% In traditional signing, the state of signing is ...

\subsubsection{Open Source}

\begin{figure}
	\centering
	\includegraphics[width=1\linewidth]{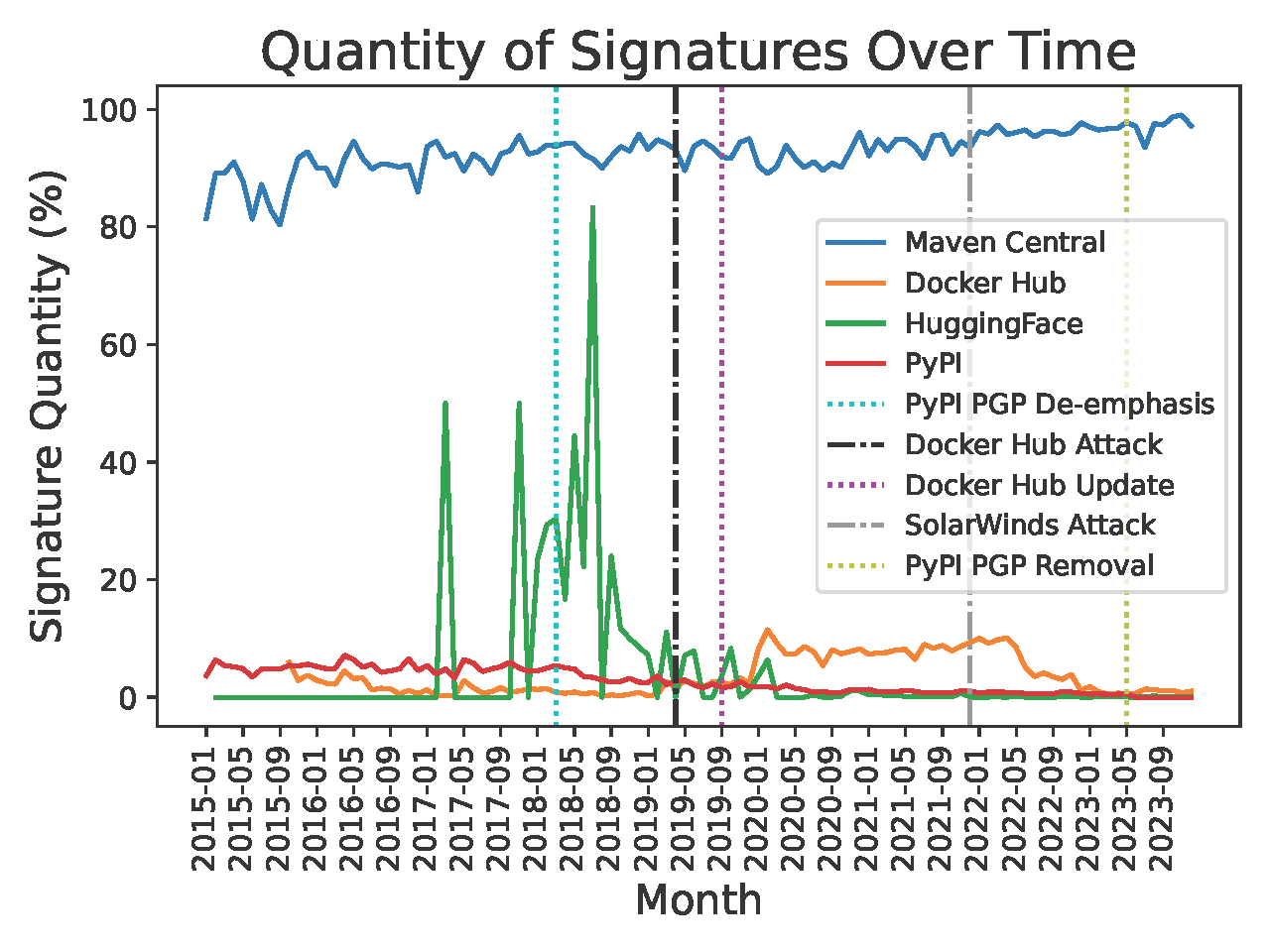}
	\caption{
		Software signing trends in four public software package registries.
		Figure is from Schorlemmer \etal~\cite{schorlemmer_signing}.
	}
	\label{fig:traditional_trend}
\end{figure}

In open source, the state of signing is generally poor.
This is depicted in \cref{fig:traditional_trend}.
The number of user created signatures on high profile platforms like Docker Hub, PyPI, and Hugging Face has historically been low.
There are many potential reasons for this, but a simple explanation is that platforms do not require signatures~\cite{schorlemmer_signing}.
This is particularly evident when comparing Maven Central (a platform that requires signing) to the rest in \cref{fig:traditional_trend}.

Regardless of the reason, the small amount of signing on these platforms decreases the effectiveness of signing.
Since so few people sign their projects, verifying signatures on a project that relies on several dependencies is very hard to do.
Furthermore, this lack of verification dissuades maintainers from going through the hassle of creating and managing signatures.
PyPI is a prime example of this, after years of poor adoption, the registry maintainers decided to stop supporting PGP.
In these circumstances, establishing provenance is hard.

The case is a bit different in ecosystems that require signing.
In instances like the Debian project or on platforms like Maven Central, contributors must sign and maintain keys.
This means that other users of those ecosystems can verify provenance.
Unfortunately, it also means that users of these platforms have to deal with the drawbacks of traditional signing schemes.

\subsubsection{Commercial Software}

\iffalse
\myquote{We sign our code and then
	we also verify that it was built in our specific build system}

\myquote{I think everybody
	on the team does sign their commits, but we don’t really
	verify it...If somebody stopped signing their commits, I don’t
	think much would really happen}
 \fi
 
% In commercial software, the adoption and use of signing is more profound. Kalu \etals ~\cite{} recent study shows that all interviewed organizations used signing in some or all parts of their software process. While the tooling choice, perception of importance, and experienced challenges vary, the use of signing as a technique was still highlighted as popular and and in use. 
In commercial software, the implementation and utilization of signing are significantly more pronounced.
% Recent surveys~\cite{WheelerSurvey,LadisaSOK} indicate that organizations valued signing.
% However, a general issue of the cost of setting up signing infrastructure.
Recent surveys ~\cite{WheelerSurvey,LadisaSOK} indicate that organizations value signing (and, therefore, provenance). However, the cost of setting up a signing infrastructure is a general concern.
Although there is a diversity in the choice of tools, the perceived importance, and the challenges encountered, the adoption of signing as a practice remains widely recognized and actively employed.

%\subsection{Historical Trends}
%Some prose about how signing has been used in the past.

\section{NEXT-GEN. SOFTWARE SIGNING}

%\myquote{There are teams who've been doing this for quite a while...Selling [a new approach] is a big challenge, right? Because we need to prove that it is much more secure than what has been there in place.}{Manager} % Kalu S18

\myquote{Sigstore comes in...keyless signing...you don’t have to worry about long-lasting static SSH keys, or keys being compromised, or rotating keys...you don’t have big servers to store all these signatures}{Manager} % Kalu S18

\myquote{[Sigstore’s] other strength...I can associate my identity with an OIDC identity as opposed to necessarily needing to generate a key and keep track of that key and yada, yada. So that’s super useful because I could say, `Oh, this is signed with [a] GitHub identity.' So unless my GitHub identity has been compromised, that’s much better.”}{Member of senior management} % Kalu S17

In the past several years, many structural improvements have been made to signing schemes.
We call the result \textit{next-generation} signing tools.

\subsection{Changes to Signing}
Next-generation signing tools are motivated by the challenges associated with traditional signing methods, such as key management, identity verification, and transparency.
These tools still use some of the primitive constructs (\eg cryptographic algorithms) of traditional signing.
However, they change the signing workflow depicted in \cref{fig:trad-vs-next-gen-signing} to address shortcomings in traditional signing tools.
% Next-generation signing tools are motivated by the major problems of traditional signing --- key management, identity verification, and transparency. While Next-generation signing tools still operate on teh principles outlayed in ~\cref{fig:traditional_signing}, addon features to resolve the issues of traditional signing have been built in them to resolve the issues mentioned above.
% Next generation signing tools address these issues in several ways.

% The three primary issues with traditional signing are key management, identity verification, and transparency..

\subsubsection{Key Management}
One way to address key management issues is to use ephemeral keys.
Ephemeral keys are temporary keys generated for a single signing session and discarded afterward.
They are created using secure key generation algorithms, with the private key securely generated, used for signing, and discarded after a single use within a validity window.
The corresponding public key is then bound to the signer’s verified identity, ensuring authorship, and made available for signature verification.
These keys do not need to be stored and managed by a user.
This reduces the burden on users to ensure that they follow best practices for keys over their lifetime.

\subsubsection{Identity Verification}
Linking a public key to an identity is a difficult problem.
Unlike traditional signing methods that rely on public key infrastructure, the web of trust model, or even off hand verification (\eg by meeting with the keyholder in person during a conference), next-generation signing tools leverage Single Sign-On (SSO) protocols like OpenID Connect (OIDC), SAML, and OAuth.
Next-generation signatures bind ephemeral public keys to an account managed by an identity provider.
For example, using the OAuth 2.0 protocol, a signing system might verify that a user has access to an account through an identity provider (\eg{Google}).
After verifying that a user has access to an account, the signing system would bind an ephemeral public key (\ie, issue a certificate) to that account.
This process links each signature to a verified identity, ensuring that it can be confidently traced back to the signer, thereby strengthening both security and trust.

\subsubsection{Shift Towards Key and Software Transparency}
Transparency allows verifiers to ensure that key-issuers do not misbehave (e.g., by handing out a key to another party).
Beyond allowing for easy key management for users, next-generation signing systems integrate transparency for keys issued and signatures created.
This transparency most often takes the form of a public append-only logs for signatures and identity-key bindings.
%Transparency logs are a way to ensure that all signatures are publicly available.
This allows users to actively monitor signatures and ensure that they are not being withheld or re-used.
%Furthermore, transparency logs can be used to ensure that all signatures are publicly available since they are stored in a public log.

\subsection{How Next-Gen. Signing Works}

Compiling these changes to signing, we can see the general form of next-generation signing tools.
The right half of \cref{fig:trad-vs-next-gen-signing} shows an example protocol for next-generation software signing.

\subsubsection{Components}
There are several components in a next-generation signing system.
Generally, the following services are used:
\begin{itemize}
    \item An \textit{identity provider} verifies that users have access to an account.
    \item A \textit{certificate authority (CA)} acts as a trust-root and binds public keys to accounts verified by identity providers.
    It issues a certificate to verify the binding.
    \item A \textit{certificate log} keeps a public, append-only copy of certificates issued by the CA to different accounts.
    This allows an account holder to verify that only authorized certificates have been issued.
    \item A \textit{signature log} keeps a public, append-only copy of signatures and their associated certificates.
    This allows verifiers to ensure that signatures were created during the lifespan of ephemeral keys and that the public key is valid.
\end{itemize}
Each of these services runs independently.
As a result, users can check for the compromise of one component by auditing the others.
For example, if an attacker attempted to modify an entry in the signature log, users could detect that either the certificate had changed (\ie, by referencing the certificate log) or that the signature had changed (\ie, by checking the signature with the public key embedded in the certificate).

\subsubsection{Workflow}
First, an author creates a package (A) and chooses one of the next-gen signing tools.
The tool then follows a supported authentication protocol to redirect the signer to an identity provider of their choice (B). After successful authentication, the identity provider issues an identity token to the signing tool (C). 
This token serves as non-repudiable proof that the user authenticated successfully. The tool then generates an ephemeral key pair (D).

The signer's identity, along with the public key, is bound together through a short validity certificate issued by a Certificate Authority (CA), or alternatively, it may be bound directly within the issued token (E).
While the specific method of binding may vary, the main goal is to link the public key to an identity.
If a certificate is used, it is published to a public log (F).
Once the identity binding is complete, the private key is used to sign the package during the short lifetime of the certificate (G), creating a signature whose authorship is verifiable.

To verify the software, a user retrieves the signed software, along with the certificate and the expected signer’s identity.
First, the certificate is verified against the root CA and key chain (H).
Next, the certificate is checked against the certificate transparency log (I).
The bound identity in the certificate is then compared with the expected identity.
If all checks pass, the public key is retrieved from the certificate and used to verify the signature on the package.
If all validations are successful, the user has confidence in the origin of the package.

\subsection{Examples of Next-Gen. Signing Platforms}

%\JD{One paragraph about sigstore}
Various systems have been engineered to address the challenges of traditional signing in the last couple of years.
Perhaps one of the most visible ones is the Sigstore project, which simplifies signature creation and verification.
Sigstore is an open source and free-to-use project that is supported by the Linux Foundation and several other organizations.
In order to achieve this, Sigstore leverages existing standards such as OIDC for identity verification, as well as Transparency Logs for software and key transparency.

Similarly to Sigstore, the Open PubKey (OPK) project provides a similar construction (identity providers to issue single-use ephemeral keys).
However, it tries to minimize infrastructure requirements. 
Even though OPK is designed to be applicable to other use cases (\eg, ssh authentication using identity providers and end-to-end encrypted chat), it is also applicable for next-generation software signing.
Currently, OPK is used by DockerHub to sign their official images~\cite{DockerHubUsingOPK}. 
A generalization of this construction is also being developed by standards organizations.
Of particular interest, the IETF's Proof of Issuer Key Authority (PIKA) can be applied to the same usecases as Sigstore and OPK~\cite{PIKA}.

\subsection{Challenges}
Although next-generation signing tools address many shortcomings of traditional signing methods, some challenges remain.
Even when using next-generation signing tools, users must still decide who to trust, watch for anomalous behaviors, and decide what to do when signature checks fail.

\subsubsection{Identity Verification}
Next-generation signing techniques rely on identity providers (\eg Google and GitHub) to bind accounts with ephemeral public keys, but users must still decide which identities to trust.
Signatures only indicate where the software came from, not if the producer is trustworthy.
If users decide to trust a bad actor, they might receive signed, but malicious, software artifacts.

\subsubsection{Transparency}
Next-generation signing still requires active analysis on the behalf of its users.
Tools like transparency logs make this job easier, but users still need to watch for malicious behavior.
For example, if identity providers are compromised (\eg leaked/stolen credentials), then malicious signatures can be created.
This requires the signer to actively monitor certificate and signature logs for unauthorized behavior and continuously check for account breaches.

\subsubsection{Signature Checks}
Some software artifacts may still not have signatures even if next-generation signing is available.
This is already a problem in current package registries where signing is not required~\cite{schorlemmer_signing}.
Unfortunately, this leaves the question: What to do with unsigned dependencies?
Furthermore: What to do with bad signatures?
Without widespread adoption in an ecosystem, the value of signing is greatly depreciated; and without a security policy for signatures, any benefit they provide may go unnoticed.

\subsection{Current State}

Next-generation software signing solutions have experienced a significant rise in popularity since their introduction.
Industry (\eg companies like Shopify, Autodesk and Verizon) and Open Source (\eg the Python interpreter and Kubernetes) organizations alike have integrated these schemes to establish provenance~\cite{SigstorePaper, SperanzaPaper}.
These successes have led to public recognition from government strategies such as the US government CISA's ``Improving Security of Open Source Software in Operational Technology and Industrial Control Systems'' as well as industry organizations such as JPMorgan's CISO. 
However, concrete measurements of adoption of these technologies are lacking, and thus it is difficult to fully assess whether this publicity is affecting adoption.
%\JD{Santiago, can you cite Sigstore~\cite{SigstorePaper} and Speranza~\cite{SperanzaPaper} papers for adoption info? Also refer White House guidance and corporate contributors/boards. Then say ``For concrete measurements, there is not data, so we got some.''}

% goals/consturcts
As a step toward quantifying the adoption status of next-generation signing tools, we looked at trends related to one of these tools, Sigstore.
Our goal was to estimate its usage.
We examined two aspects of usage: download rates of the Sigstore toolkit, and analyzing use and discussion of the tools and associated techniques, such as npm's provenance feature,\footnote{See \url{https://docs.npmjs.com/generating-provenance-statements}.} which use Sigstore under the hood.

\begin{figure}
	\includegraphics[width=\linewidth,page=3]{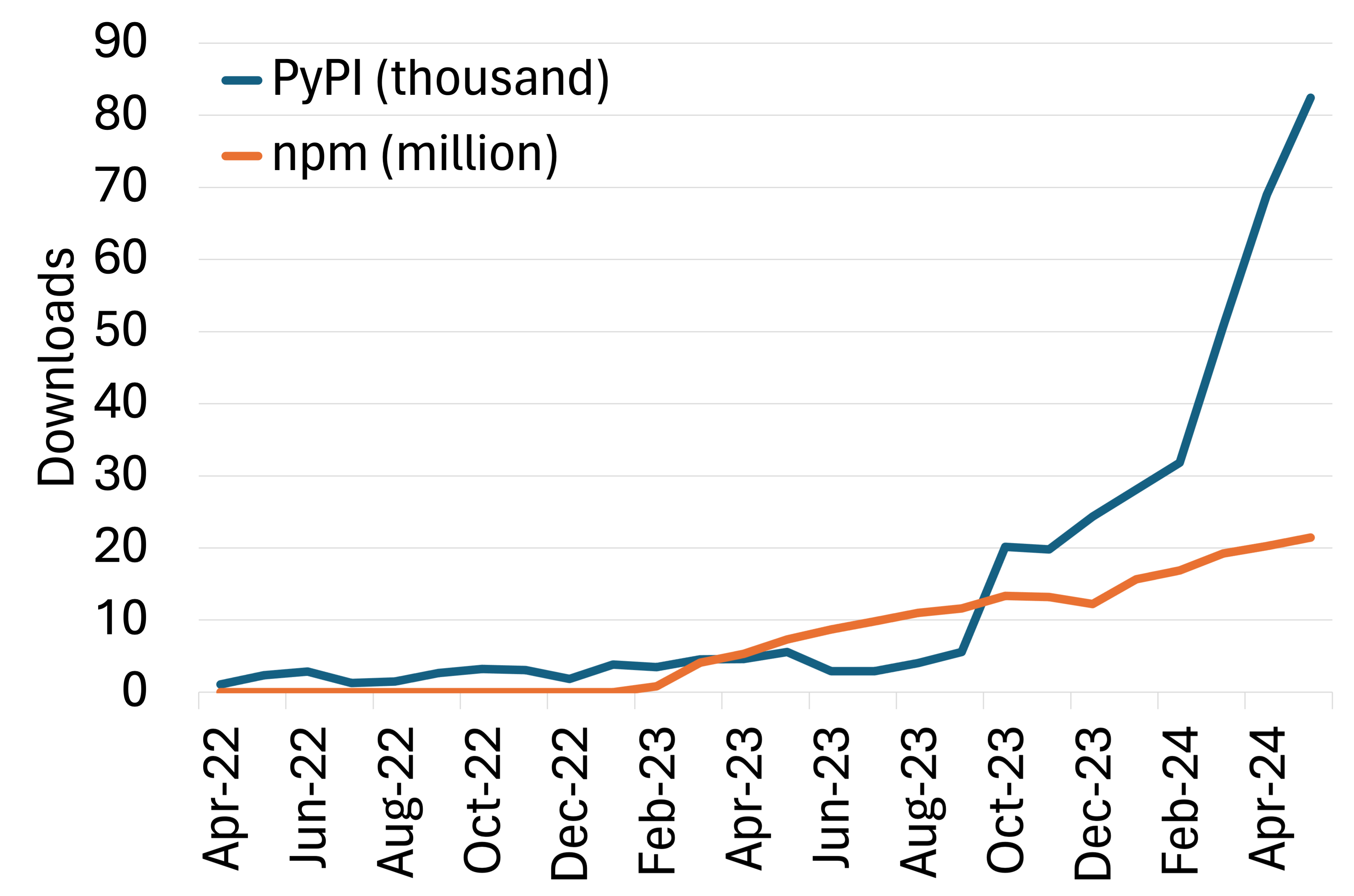}
	\caption{
      The Sigstore project has tooling available in several programming languages.
      This chart shows the downloads over time of the associated package for the Python-PyPI package (blue, in thousands) and the JavaScript-NPM package (orange, in millions).
	}
	\label{fig:next_gen_trend}
\end{figure}

% methods/results
% To analyze the rate at which next-generation signing tools are downloaded, we collected download rates for some Sigstore toolings over the past 24 months.
To estimate the adoption rate of next-generation signing tools, we obtained the download statistics for various Sigstore-related tooling over the past 24 months. 
Two Sigstore tools were checked: NPM and PyPI. The PyPI download data was collected using the PyPI database as queried through Google's BigQuery. The NPM download data was collected using npm-stat.
\cref{fig:next_gen_trend} summarizes our result.
%The download rates for both Sigstore distributions can be seen in \cref{fig:next_gen_trend}.
Both PyPI and npm versions of the tooling are seeing increasing use, measured in the tens of thousands to millions of downloads over time. %, and suggest that it may soon replace the traditional methods still being used today.

To observe the usage and discussion of next-generation signing tools, we used Sourcegraph's Code Search to query GitHub repositories for keywords relating to the tools.
These keywords were determined based on package names (top half of the table) and indicators of provenance workflows within repositories (bottom half of the table).
The results of these keyword searches are shown in~\cref{table:next_gen_mentions}.
%The results shown imply the scale at which provenance and next generation signing tools are implemented and discussed. 

% takeaways
%We can see through a positive download rate from a next generation signing tool that these adoption of these tools is on the rise.
%Furthermore, the substantial indication of usage and discussion reinforces this belief. 

 \begin{table}
 	\vspace*{4pt}
	\caption{Estimates of use of next-gen signing tools via keyword searches of open-source repository files.
		% \textbf{Measurement:} specifies the variable meassured for each listed Next-generation signing tool --- \textbf{\textit{MS}} Mentions of specified signing tool name, \textbf{\textit{NR}} Number of repositories using specified tool
		%\textbf{\textit{Method:}} Describes the method employed in the measurement. 
  %\textbf{\textit{MK}} -- Method measures the number of repositories containing the specified tool keyword, \textbf{\textit{MID}} -- Measured the number of “id-token” fields set to write in GitHub workflow files (as this is necessary for using the specified tool), \textbf{\textit{MPW}} -- Measured the number of repositories using the ``--provenance'' flag in their workflow files
  The Method was as follows:
    For the tools in the top half of the table, we searched for the tool name as a string, anywhere in the project.
    For the tools in the lower half of the table, we check flags or fields in GitHub workflow files that indicate the use of Sigstore.
    In npm provenance workflows, we check if the ``id-token'' field is set to write.
    For npm publishing with provenance, we checked for the ``--provenance'' flag.
	}
 	\label{table}
 	\tablefont
 	\begin{tabular*}{17.5pc}{@{}p{75pt}p{30pt}<{\raggedright}p{35pt}<{\raggedright}p{35pt}<{\raggedright}@{}}
 		\toprule
		\textbf{Approach}                        & \textbf{Measure} & \textbf{Mentions} & \textbf{Repos} \\
 		\midrule
		sigstore                             & Name               & 143,830           & 5,229          \\[3pt]
		sigstore-python                      & Name & 1,041             & 396            \\[3pt]
		sigstore-java                        & Name & 194               & 21             \\[3pt]
		sigstore-rs                          & Name & 235               & 23             \\[3pt]
  		in-toto                              & Name & 68,752            & 985            \\[3pt]
            \midrule
            npm provenance workflow              & GitHub Workflow              & 37,363            & 26,255         \\[3pt]
		npm repos publishing with provenance & GitHub Workflow & 1,427             & 896            \\[3pt]
 		\botrule
 	\end{tabular*}\vspace*{8pt}
	\label{table:next_gen_mentions}
 \end{table}
 
Both measures (\cref{fig:next_gen_trend}, \cref{table:next_gen_mentions}) indicate increasing interest in Sigstore, the next-generation signing tool that we examined.

\section{CONCLUSIONS}

Software supply chains are not new, but they are more visible than ever before.
Software signing --- of code, of attestations, of bills of materials, and of all other aspects of consuming a software component --- is a key technique in ensuring that supply chain artifacts have not been modified since they were authored by trusted parties.
Although the previous generation of signing tools left much to be desired, the next generation of tools has addressed many of those challenges and is seeing widespread adoption.

The technological capability is now present for any component provider to easily sign their products.
Software signature creation and verification are already commonplace.
While selecting components, software engineers can and should now rely only on components bearing the signature of the author.
%The perspective of software consumers can shift from a passive need to change from a passive role 
Conversely, software producers should expect that their consumers will begin to demand these signatures, as a result of due diligence or the need to meet current or anticipated regulations regarding software provenance.
Soon, checking signatures before coding will be a normal part of component selection.

Finally, we caution the reader that provenance --- even cryptographically-assured provenance using signatures --- does not guarantee correctness nor security.
Provenance information reduces the risk of certain classes of attacks, such as man-in-the-middle substitutions.
Provenance information also improves the engineer's ability to estimate the quality of the components they rely on, \eg based on the reputation of the supplier.
But guaranteeing the correctness and security of a component (not to mention the resulting system) is a task for formal methods, which is another endeavor for computing.

%\myquote{I'm not confident that...customers actually ever verify those signatures}{Manager} % Kalu S7

%\myquote{Most software package ecosystems that I consume packages from do not support signing consistently enough to put much stock in that}{Manager} % Kalu S13

%\myquote{My external, looking-in perspective of things, is that there's a mix of anticipation of regulation, as well as reaction to regulation...[T]here's a lot of trying to predict what is going to come, what the government is seeking for as far as tech regulation, and trying to get ahead of that curve}{Technical leader} % Kalu S8

\vspace*{-8pt}
\section{ACKNOWLEDGMENTS}
The authors thank
  Zach Steindler of GitHub for his feedback on our estimates of npm provenance adoption,
  and
  Chinenye Okafor for her input on the design of next-gen signing systems.
This work was supported by the US National Science Foundation under award \#2229703, and by funds from Google and Cisco.

\def\refname{REFERENCES}

\vspace*{-8pt}

 \begin{IEEEbiography}{Taylor R. Schorlemmer}{\,}
 is a Lieutenant in the US Army.
 His research interests include software engineering and cybersecurity, with an emphasis on software signing and software supply chains.
 Schorlemmer received his MSc in Computer Engineering from Purdue University.
 He is a Member of the IEEE.
 Contact him at tschorle@purdue.edu.
 \end{IEEEbiography}

 \begin{IEEEbiography}{Ethan H. Burmane}{\,} is an undergraduate student in the Elmore Family School of Electrical \& Computer Engineering at Purdue University, at West Lafayette, IN, 47907, USA.
 His research interests include software engineering and cybersecurity.
 Contact him at eburmane@purdue.edu.
 \end{IEEEbiography}

 \begin{IEEEbiography}{Kelechi G. Kalu} {\,}
 is a PhD student in the Elmore Family School of Electrical \& Computer Engineering at Purdue University, at West Lafayette, IN, 47907, USA.
 His research interests include software engineering and cybersecurity, with a focus on software supply chains.
 Kalu received his BSc in Electronic and Computer Engineering from Nnamdi Azikiwe University, Nigeria.
 Kalu is a Member of the IEEE.
 Contact him at kalu@purdue.edu.
 \end{IEEEbiography}
 
 \begin{IEEEbiography}{Santiago Torres-Arias} {\,}
 is an assistant professor in the Elmore Family School of Electrical \& Computer Engineering at Purdue University, at West Lafayette, IN, 47907, USA.
 His research interests include cybersecurity, cryptography, and computer systems, with an emphasis in software supply chain security.
 Torres-Arias received his PhD in Computer Science from New York University (NYU).
 He is a Member of the ACM and a Member of the IEEE.
 Contact him at santiagotorres@purdue.edu.
 \end{IEEEbiography}
 
 \begin{IEEEbiography}{James C. Davis} {\,}
 is an assistant professor in the Elmore Family School of Electrical \& Computer Engineering at Purdue University, at West Lafayette, IN, 47907, USA.
 His research interests include human and technical aspects of software engineering and cybersecurity, with an emphasis on the analysis of failures.
 Davis received his PhD in Computer Science from Virginia Tech.
 He is a Member of the ACM and a Senior Member of the IEEE.
 Contact him at davisjam@purdue.edu.
\vadjust{\vfill\pagebreak}
 \end{IEEEbiography}

\end{document}